# TEMPERATURE DEPENDENCE OF THE ELECTRICAL TRANSPORT PROPERTIES OF MULTIPLAYER GRAPHENE

E. S. Sadki [1], H. Okazaki [2], T. Watanabe [2], T. Yamaguchi [2], Y. Takano [2]

[1] UAE University
Abu Dhabi, United Arab Emirats
e_sadki@uaeu.ac.ae
[2] National Institute for Materials Science
Tsukuba, Japan

## 1. Introduction

Graphene has very interesting fundamental properties and promising applications [1]. Consequently, this material also re-generated an interest in studying its multilayer systems, ranging from few to large numbers of graphene materials (i.e. graphite) [2]. However, the low temperature properties, and superconductivity in particular, have not been sufficiently investigated in graphene and few-layer graphene systems. This is despite the several experimental demonstrations of superconductivity in intercalated graphite (e.g. [3]), and few theoretical predictions on superconductivity in graphene and few-layer graphene (e.g. [4]). Some groups have reported superconductivity in multilayer graphene induced by electrical gating [5] and potassium doping [6]. However, these results have not been confirmed by others so far. In this work, multilayer graphene (MLG) systems with different numbers of graphene layers are fabricated, and their electrical transport properties are measured at low temperatures to explore some of the above theoretical and experimental reports.

## 2. Experimental

The MLG thin films samples are prepared by scotch tape method of kish graphite onto a 300 nm thermally grown silicon oxide on silicon wafer ($SiO_2$ / Si) [7]. The MLGs of interest are initially located under an optical microscope, patterned by electron beam lithography, followed by gold / titanium (50 nm / 5 nm) evaporation, and finally lifted-off, to form electrodes configuration for electrical transport characterization. **Figures 1a** and **b** show an example of six metal terminals deposited on an MLG sample. An atomic force microscope (AFM) (Nanocute, SII NanoTechnology) is operated in AC-mode to measure the thickness of the MLG samples.

**Figure 1c** shows an AFM topography image of an MLG sample, with its measured thickness. Raman spectroscopy (RAMAN–11, Nanophoton), equipped with a laser of 532 nm in wavelength, is used to estimate the number of layers in the MLG samples. The electrical resistance measurements of the samples are conducted in a four-terminal configuration from room temperature down to 2 K, using a physical properties measurement system (PPMS, Quantum Design) with a constant current of typically 100 nA. For comparison, the same electrical measurements were also conducted using an AC lock-in amplifier technique.

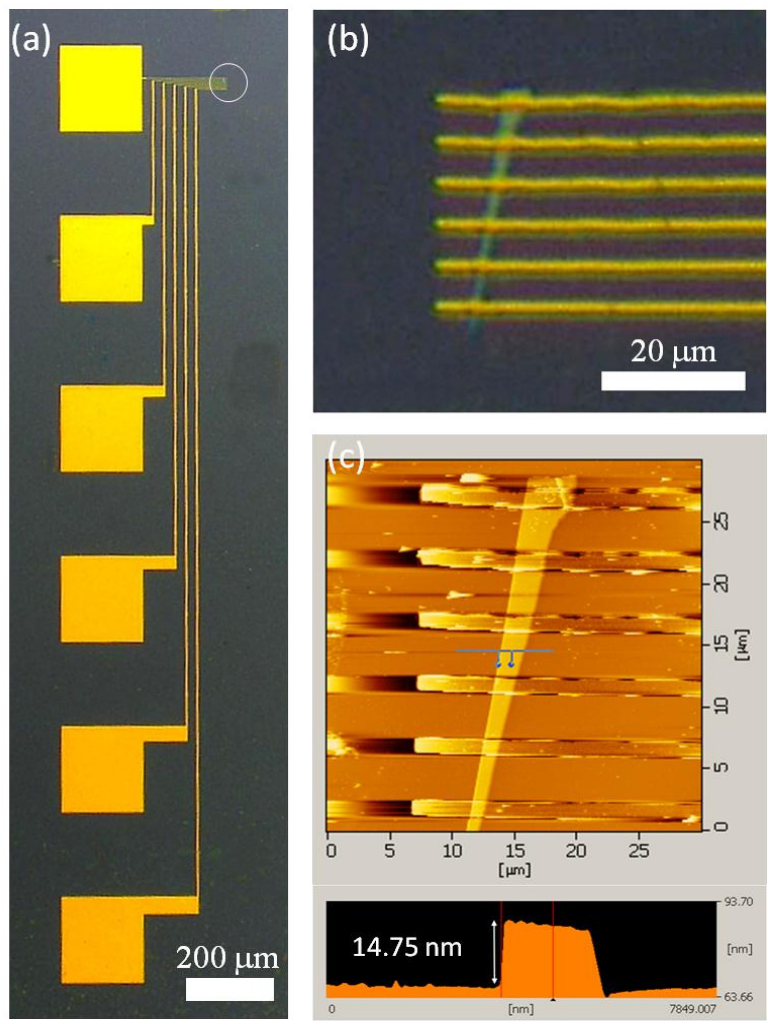

**Figure 1.** Optical microscopy photographs of (a) the fabricated gold electrodes configuration, with a MLG sample located in the area within the circular indicator, and (b) a close-up on MLG1 sample under the gold terminals; (c) an atomic force microscopy (AFM) topography image of MLG1, with the inset below showing the measured thickness of the sample.

### 3. Results and discussion

The AFM-measured thickness of three samples MLG1, MLG2, and MLG3, are 14.75, 3.93, and 3.37 nm, respectively. Assuming a thickness of 0.345 nm for a single layer graphene, the estimated number of graphene layers is 42, 8, and 3, for MLG1, MLG2, and MLG3, respectively. These figures are compared with Raman spectroscopy data below.

**Figure 2** shows the Raman spectrum of the samples. The G-band and 2D-band peaks are clearly observed (2D-band was not measured for MLG1). These peaks are used to obtain many structural and physical properties of monolayer graphene and MLGs [8]. In particular, these peaks were used to estimate the number of graphene layers in MLGs [9]. For example, the number of graphene layers was correlated to both the intensities of the G-band peak and 2D-band peak, as well as to their positions [10 – 12].

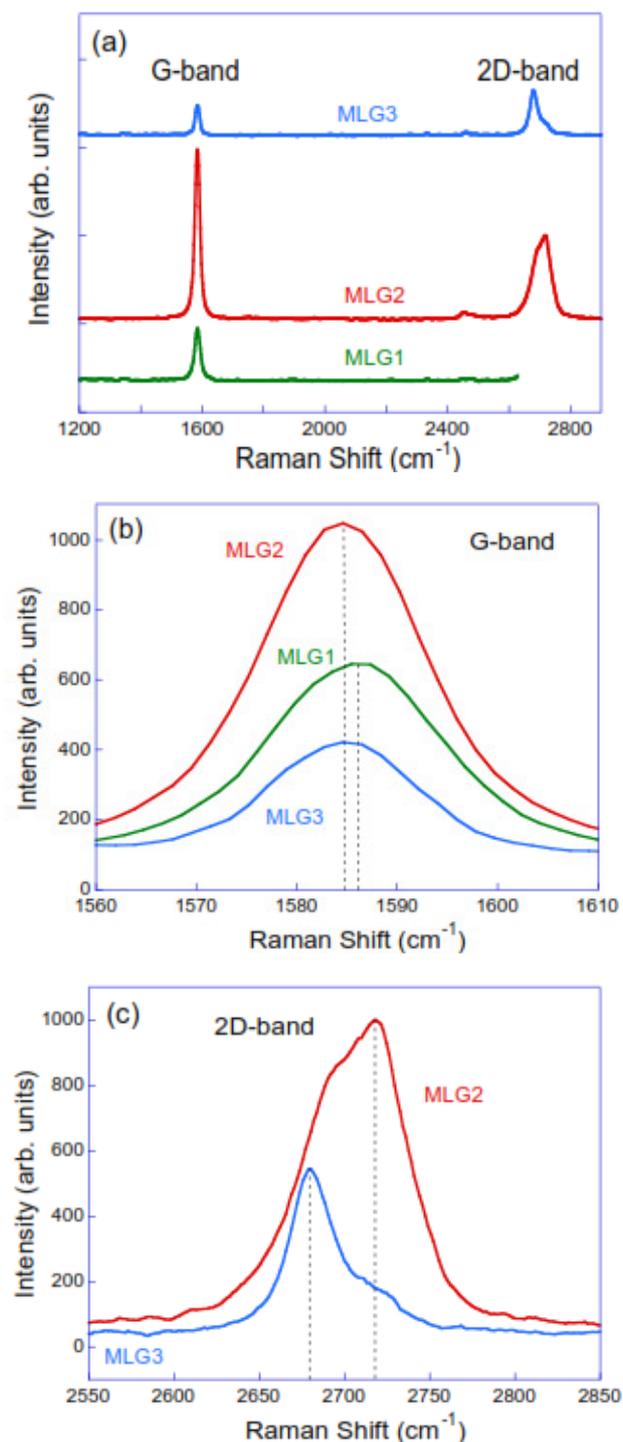

**Figure 2.** (a) Raman spectrum of the MLG samples excited by a 532 nm laser. Raman spectrum (b) around the G-band, and (c) the 2D-band. Dotted vertical lines show the positions of the respective peaks.

As shown in **Figure 2b**, the G-band peaks of sample MLG1 is located at 1586.4 cm$^{-1}$, and that of both MLG2 and MLG3 coincide at the same position 1584.6 cm$^{-1}$. In general the frequency is expected to shift lower with increasing the number of layers [8]. However, for MLGs with less than around 10 layers, the change is not systematic [11, 12]. Therefore, the peaks positions of the G-band of our samples cannot be used to accurately determine the number of layers. The ratio of the intensities of the G-band peaks of MLG2 on MLG3 is about 2.3 (see **Table 1**). This value is in good agreement with the data from [11, 12] that corresponds to a number of layers of 7 and 3 for MLG2 and MLG3, respectively. This is in good agreement with the AFM data. However, errors are expected in AFM measurements on MLGs due to their dependence on the operating conditions [13], and hence, Raman data is preferable for this calculation as the difference in G-band peak intensities significantly changes with the number of layers. It is noted that the intensity of the G-band peak of MLG1 is lower than that of MLG2. This is expected for graphene layers of more than about 7 to 9 [11, 12].

**Table 1.** Summary of samples properties.

| Sample | AFM thickness, nm | G-band position, cm$^{-1}$ | G-band peak intensity, a.u. | 2D-band position, cm$^{-1}$ | 2D-band peak intensity, a.u. | Number of graphene layers |
|---|---|---|---|---|---|---|
| MLG1 | 14.75 | 1586.4 | 566 | – | – | 42 |
| MLG2 | 2.93 | 1584.6 | 789 | 2718 | 924 | 7 |
| MLG3 | 1.28 | 1584.6 | 335 | 2680 | 505 | 3 |

The 2D-band peaks for MLG2 and MLG3 are shown in **Figure 2c**. The peaks positions are located at 2718 and 268 cm$^{-1}$ for MLG2 and MLG3, respectively. This shift of the 2-band peak to higher frequencies with increasing number of graphene layers is in agreement with both experiments and theoretical predictions [8]. By comparison with the data from [12], the general shape of our peaks corresponds indeed to the 3 and 7 layers graphene systems. All the above data is summarized in **Table 1**. **Figure 3** presents the electrical resistance versus temperature results. With decreasing temperature, all MLG samples show first an increase in the resistance up to a maximum point, then a decrease, and finally a small increase at the lowest temperatures. The increase in the resistance up to the maximum point was more dramatic with increasing the number of graphene layers in the samples, with normalized resistance to the 300 K value (i.e. $R/R_{300 K}$) of 1.8, 1.2, and 1.1, for MLG1, MLG2, and MLG3, respectively. The position of the maximum value of the resistance is located at 19, 50, and 39 K, respectively.

The above results are quite different from that of bulk graphite that shows a monotonic decrease in the resistance with decreasing temperature, i.e. metallic-like behavior [14]. However, similar qualitative effects to our data were observed when the thickness of the graphite samples was reduced to below 20 mm [15, 16]. The thinnest measured samples from these groups were 12 [15] and 13 nm [16]. The thickness of these samples correspond roughly to our MLG1, and indeed the value of the normalized resistance at the peak value (~ 1.3), and its position (~ 50 K) from [15] are in reasonable agreement with our data for MLG1. However, for the 13 nm thickness sample from [16], no peak was observed. Nevertheless, the same group [16] observed a peak with $R/R_{300 K}$ ~ 1.5 at 35 K for a 20 nm thick sample, which is strikingly very similar to MLG1. This peak was explained by competing contributions between a semiconducting-like behaviour of the intrinsic graphene layers and metallic-like behaviour originating from interfaces or defects between these layers [16], or by assuming the simple two band model (STB) of overlapping between the electrons and holes bands around the Fermi level of graphite [15].

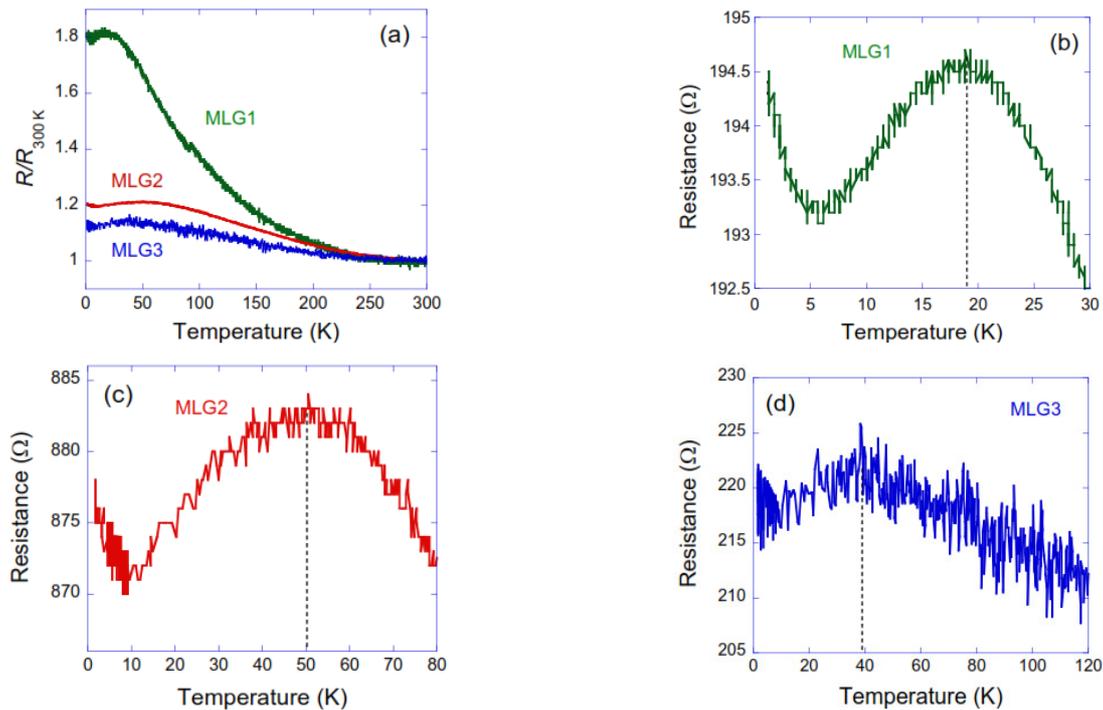

**Figure 3.** (a) Normalized resistance ($R / R_{300\,K}$) of the MLG samples measured from 300 down to 1.5 K. The resistance versus temperature of the samples (b) MLG1, (c) MLG2, and (d) MLG3, measured before the maximum resistance and down to 1.5 K. The dotted lines show the location of the onset of the observed down-turn of the resistance.

The peaks in the resistances of MLG2 and MLG3 observed are less pronounced than that of MLG1, with values of about 20 and 10 % above $R_{300\,K}$, for MLG2 and MLG3, respectively. This is inconsistent with the experimental data from [15] and [16] that predict that the maximum normalized resistance should increase with reducing the number of graphene layers. However, these groups never measured MLG samples as thin as ours. It is also noted that the signal to noise ratio decreases with decreasing the number of graphene layers. This could be explained by increased sensitivity to the environment for thinner samples, especially MLG3.

In order, to explore all the above theories and others, further structural analysis on defects / interfaces in our samples, as well as electrical transport measurements under magnetic fields are required, which is beyond the scope of this paper. Finally, it is noted that superconductivity, which is characterized by a dramatic drop of the electrical resistance to zero, is not observed in our MLG samples. A way to explore this possibility is by controlling the electronic carrier concentration by electronic gating [5], ionic-liquid doping [17], or intercalation [3]. These could be very interesting paths to explore in the future.

## 4. Conclusion

In conclusion, different multilayer graphene (MLG) samples were fabricated, and the number of their graphene layers were determined from both Atomic Force Microscopy (AFM) and Raman Spectroscopy analysis. The electrical resistance is measured from room temperature down to 2 K. The measured electrical resistance of the MLGs shows an increase with decreasing temperature, and then a downturn decrease at lower temperatures. The increase in the resistance up to the maximum point was more dramatic with increasing the number of graphene layers in the samples. Superconductivity is not observed in these samples.